\documentclass[english,aps,prper,longbibliography,reprint,superscriptaddress]{revtex4-1}   
\usepackage{lipsum}
\usepackage[T1]{fontenc}	
\usepackage[latin9]{inputenc}	
\usepackage{geometry}		
\usepackage{graphicx}
\usepackage{color}
\usepackage[above,below]{placeins}	
\usepackage{times}
\usepackage[colorlinks=true,allcolors=blue]{hyperref}
\usepackage{multirow}
\usepackage{rotate}

\usepackage{lineno,xcolor}
\usepackage{amsmath,amssymb}
\usepackage{mathrsfs}
\usepackage{array}
\usepackage[caption = false]{subfig}
\usepackage{makecell}

\begin{document}
\title{A framework for the natures of negativity in introductory physics} 

\author{Suzanne White Brahmia}
\affiliation{Department of Physics, University of Washington, Box 351560, Seattle, WA 98195-1560, USA}

\author{Alexis Olsho}
\affiliation{Department of Physics, University of Washington, Box 351560, Seattle, WA 98195-1560, USA}

\author{Trevor I.\ Smith}
\affiliation{Department of Physics \& Astronomy and Department of STEAM Education, Rowan University, 201 Mullica Hill Rd., Glassboro, NJ 08028, USA}

\author{Andrew Boudreaux}
\affiliation{Department of Physics \& Astronomy, Western Washington University, 516 High St., Bellingham, WA 98225, USA}

\begin{abstract}
Mathematical reasoning skills are a desired outcome of many introductory physics courses, particularly calculus-based physics courses. Positive and negative quantities are ubiquitous in physics, and the sign carries important and varied meanings. Novices can struggle to understand the many roles signed numbers play in physics contexts, and recent evidence shows that unresolved struggle can carry over to subsequent physics courses. The mathematics education research literature documents the cognitive challenge of conceptualizing negative numbers as mathematical objects---both for experts, historically, and for novices as they learn. We contribute to the small but growing body of research in physics contexts that examines student reasoning about signed quantities and reasoning about the use and interpretation of signs in mathematical models. In this paper we present a framework for categorizing various meanings and interpretations of the negative sign in physics contexts, inspired by established work in  algebra contexts from the mathematics education research community. Such a framework can support innovation that can catalyze deeper mathematical conceptualizations of signed quantities in the introductory courses and beyond. 
\end{abstract}

\maketitle

\section{Introduction}
\label{sec:intro}

\subsection{Motivation}
\label{ssec:motivation}
Experts in physics translate fluidly between different representations of phenomena. To an expert, a physics equation ``tells the story'' of an interaction or process. For example, when reading the equation \linebreak $x(t)=+40~\textrm{m}+(-5~\textrm{m/s})t+\frac{1}{2}(-9.8~\textrm{m/s}^2)t^2$, an expert may quickly construct a mental story of the co-variation of position and time of a projectile that starts $40~\textrm{m}$ above the ground and is launched with a speed of $5~\textrm{m/s}$ vertically downward.  Part of the challenge of learning physics is developing the ability to decode symbolic representations in this manner. 

In these translation processes, experts can readily attribute specific meanings to positive and negative signs. In the example above, the positive sign in front of the term $40$ m indicates that the projectile starts at a position that is in the positive direction from the origin, \textit{upward} in this case because the gravitational acceleration is always downward and happens to be negative in this expression. The positive sign after the $40$ m term indicates that the following term, $(-5~\textrm{m/s})t$, represents an additive change in the projectile's position---a one-dimensional vector quantity that has the initial value 40~m and is in the positive direction. The sign in front of $5$~m/s indicates that the projectile is launched \textit{downward}.

Other examples of the fluid interpretation of signs abound in introductory physics:

\begin{itemize}
\item In the equation $\vec{F}_{\textrm{12}} = -\vec{F}_{\textrm{21}}$, the negative sign signals that the force exerted by object 2 on object 1 is in the exact opposite direction as the the force exerted by 1 on 2.    
\item In the expression $0-(-5~\mu$C), the first negative sign indicates that a quantity of electric charge is being removed from an electrically neutral object, while the second negative sign indicates which of the two different types of electric charge is being removed.  
\item In Faraday's law, $\mathscr{E} = -\frac{\textrm{d}\Phi_B}{\textrm{d}t}$, the negative sign reminds the expert that the voltage induced by a changing magnetic flux acts to oppose (rather than reinforce) the change that created it. 
\end{itemize}

In all of these cases, experts generally decode specific meanings of the sign quickly and effortlessly, perhaps in most cases without conscious awareness of the decoding process itself. Novices may need to spend considerable conscious effort interpreting the sign, or may fail altogether to successfully interpret it. Pitfalls likely to challenge the novice might include a tendency to overgeneralize a particular interpretation, or a lack of awareness that two nearby signs in an equation may have completely different physical interpretations. The challenge for introductory physics teaching may be compounded if the already difficult cognitive task of interpreting signs goes unaddressed when instructors themselves are not consciously aware of the mismatch between their own ability and the much lower skill level of novices. 

An additional layer of complexity is introduced by the implicit nature of the sign of many symbolically-represented quantities. For example, the expression $\Delta U_g$ is a stand-in for a generalized number of Joules, which could be positive or negative. The sign in this case would indicate whether the potential energy of some system increased or decreased during a particular process.  $U_g$ is also a stand in for a generalized number---but in contrast to $\Delta U_g$, the sign of $U_g$ tells an expert whether the energy of a system in a particular configuration is larger or smaller than the energy associated with a pre-established reference configuration.  

In a fast-paced introductory physics course, this nuanced interpretation of sign may fall by the wayside, as instructors attend to a host of other, perhaps more obvious challenges.  Unfortunately, student difficulties with decoding sign, if unaddressed, may not spontaneously resolve. In such cases, difficulties with signs could contribute to serious obstacles in the development of overall quantitative reasoning and analysis skills so highly prized by physicists.  In this article, we describe an effort to systematically parse and document the various meanings of sign in introductory physics contexts.  We hope that a taxonomy of this type can support instructor efforts to nurture their students' ability to translate between representations, and can support the efforts of physics education researchers to more fully understand the nature of student reasoning about signs and signed quantities.  
\subsection{Prior Research}
\label{ssec:prior}
Negative pure numbers represent a more cognitively difficult mathematical object than positive pure numbers do for pre-college mathematics students \cite{bishop2014}. Mathematics education researchers have isolated a variety of ``natures of negativity'' fundamental to algebraic reasoning in the context of high school algebra---the many meanings of the negative sign that must be distinguished and understood for students to develop understanding \cite{gallardo1994,thompson1988,nunes1993}. These various meanings of the negative sign form the foundation for scientific quantification, where the mathematical properties of negative numbers are well suited to represent natural processes and quantities. Physics education researchers report that a majority of students enrolled in a calculus-based physics course struggled to make meaning of positive and negative quantities in spite of completing Calculus I and more advanced courses in mathematics \cite{Brahmia2016b,Brahmia2017a}. Developing ``flexibility'' with negative numbers (the recognition and correct interpretation of the multiple meanings of the negative sign) is a known challenge in mathematics education, and there is mounting evidence that reasoning about negative quantity poses a significant hurdle for physics students at the introductory level and beyond \cite{Hayes2010,bajracharya2012,huynh2018}. 
Few published studies have focused on the use of the negative sign, i.e., \emph{negativity}, specifically in the context of the mathematics used in physics courses. Studies conducted in the context of upper-division physics courses reveal robust student difficulties associated with interpretation of the vector nature of acceleration and its representation in Newton's second law, and with contexts in E{\&}M in which there are often multiple negative signs, each with a separate meaning \cite{Hayes2010,huynh2018}.

Brahmia and Boudreaux conducted studies to probe  student difficulties with negative quantities at the introductory level. The authors constructed physics assessment items based on the natures of negativity from mathematics education research \cite{vlassis2004} and administered them to introductory physics students in the introductory sequence of courses \cite{Brahmia2016b,Brahmia2017a,Brahmia2017c}. The authors report that students struggle to reason about signed quantity in the contexts of negativity typically found in the introductory curriculum (e.g., negative work, negative direction of acceleration or electric field in one dimension), and they concluded that science contexts may overwhelm some students' conceptual facility with negativity. These studies reveal that signed quantities, and their various meanings in introductory physics, present cognitive difficulties for students that many don't reconcile before completing the introductory sequence. These difficulties then carry over into upper-division course work. 
\subsection{Contribution to the literature}
\label{ssec:contributions}
The current study advances this body of research by introducing a framework for categorizing the \emph{natures of negativity in introductory physics} (NoNIP), analogous to the natures of negativity developed in the context of algebra \cite{vlassis2004}. While we recognize that students struggle with signed quantities more generally, we choose to focus on negatively signed quantities in this work because they are the only signed quantities for which the sign is always explicitly included (e.g., a velocity in one dimension of 3 m/s is typically assumed to be in the positive direction). The intention is to provide a framework that can help researchers and instructors characterize and address the mathematical conceptualization of signed quantity in introductory physics. 

In the next section, we describe the development of the NoNIP, including its basis in an analogous framework developed by math education researchers. We present the NoNIP, along with examples of quantities and relationships that illustrate its use. We end Section \ref{sec:Natures} with a discussion of the validation of this framework in the context of introductory physics.

In Section \ref{sec:Applications}, we analyze three recent studies in upper division physics using the NoNIP framework as a lens through which student cognition can be categorized and understood. Section \ref{sec:signedQuant} describes our exploration of student understanding of ``positivity'' and the necessity of extending this work to signed quantities more generally. We discuss our conclusions and implications for instruction in Section \ref{sec:Conclusion}.

\section{Modeling the Natures of Negativity}
\label{sec:Natures}
In this section we discuss the need for and development of a framework to understand the uses of the negative sign in introductory-level physics. In section \ref{ssec:underpinnings}, we include work by researchers that has directly influenced and guided the development of the NoNIP framework. In section \ref{ssec:devNoNIP}, we describe the process by which this framework was developed, expert validated and, as a result of expert input, modified. We present the current version of the framework in section \ref{ssec:currentVersionNoNIP}.

\subsection{Underpinnings}
\label{ssec:underpinnings}

Our initial model for the natures of negativity 
in introductory physics was based on the natures of negativity in elementary algebra, described by Vlassis \cite{vlassis2004}. Vlassis summarized the work of mathematics education researchers, identifying three distinct algebraic natures of negativity. The first, referred to as the ``unary'' nature, describes situations in which a negative sign is used in close association with a single quantity, and includes the formal concept of a negative number (e.g., the number $-5$). Two additional natures signify mathematical operations: the ``binary'' nature describes various conceptualizations of the negative sign as it is used in subtraction (e.g., $5-3=2$), while the ``symmetrical'' nature describes use of the negative sign to invert (i.e., take the opposite of) a number or operation (e.g., the first negative sign in the expression $-(-5)=5$). Table \ref{tab:NoNEA} summarizes these algebraic natures. 


\begin{table*}

\caption{A map of the different uses of the negative sign in elementary algebra \cite{vlassis2004}}
\label{tab:NoNEA}
\begin{ruledtabular}
\begin{tabular}{ c  c  c }

Unary (Struct. signifier)& Symmetrical (Oper. signifier)	& Binary (Oper. signifier) \\ \hline
Subtrahend 	& \makecell{Taking opposite of \\ or inverting the operation} 		& Completing \\
Relative number		& 							& Taking away \\
Isolated number 	& 			& Difference between numbers \\
Formal concept of neg. number  	&  		& Movement on  number line
\end{tabular}
\end{ruledtabular}
\end{table*}

We found that while most uses of the negative sign that arise in introductory physics could be categorized using the map of Vlassis, the nuances of the physics described by the math were often lost. We also found ourselves tempted to represent the physical meaning of a negative sign attached to a single quantity ( ``unary'' category in Table \ref{tab:NoNEA}) using the categories intended for operations, as negative quantities in physics in some cases represent a \emph{process} rather than just an amount (e.g., system). Moreover, it was difficult for physics experts to reach consensus when attempting to categorize some quantities and relationships. The mathematical natures of the negative sign described by the Vlassis framework gave us key insights into the meaning of negative signs in the context of physics, but we ultimately concluded that the use of mathematics as a language to describe physics quantities and relationships (i.e., \emph{quantification}) requires a different categorization than pure algebra. 

Vlassis's map is based in semiotics, the study of symbols and their meanings. We believe this is appropriate for describing meanings of the negative sign in mathematics. To interpret the meaning of the negative sign in the mathematics used in physics, a blended processes framework is more appropriate. Conceptual blending theory (CBT) provides a framework for understanding the integration of mathematical and physical reasoning. In their theory, Fauconnier and Turner describe a cognitive process in which a unique mental space is formed by merging two (or more) separate mental spaces \cite{fauconnier2008}. The blended space can be thought of as a product of the input spaces, rather than a separable sum. According to CBT, development of expert quantification in physics would occur not through a simple addition of new elements (physics concepts) to an existing cognitive structure (arithmetic), but rather through the creation of a new and independent cognitive space.  This space, in which creative, quantitative analysis of physical phenomena can occur, involves a continuous interdependence of thinking about the mathematical and physical worlds. 

The design of the NoNIP was further inspired by Sherin's work on symbolic forms, which posits that 

\begin{quote}
\dots successful (physics) students learn to understand what equations say in a fundamental sense; they have a feel for expressions, and this understanding guides their work\dots from the point of view of improving instruction, it is absolutely critical to acknowledge that physics expertise involves this type of flexible and generative understanding of equations. We do students a disservice by treating conceptual understanding as separate from the use of mathematical notations \cite{sherin2001}.
\end{quote}

We take the approach that a preliminary step in helping students develop a feel for expressions in a way that can productively guide their work is to understand better how expert conceptualization is organized, and thus characterize an expert's feel for expressions in a way that is useful both to education researchers and instructors. 

We note that the physics contexts that are typically used as applications in a mathematics course are limited. Nonetheless, mathematics education research into student reasoning with quantity helps to build a framework for thinking about mathematical objects in physics contexts. In their study involving middle school through graduate mathematics students' conceptualization of negative quantity, Chiu identified three categories of metaphorical reasoning used by middle school students, undergraduate, and graduate mathematics and engineering majors during problem-solving interviews that focused on arithmetic with signed numbers. The categories isolated by the researchers are:  motion (movement along a number line), the manipulation of objects/opposing objects (removing or acting in opposition), and social transaction (associated with the experiences of giving and exchanging) \cite{chiu2001}. While these are metaphors in mathematics, they are in fact \emph{contexts} in physics in which a conceptual mathematical understanding is essential for learning the physics. In fact, the entire content of mechanics is focused on actual motion in space (not motion along an abstract number line). It is the interplay between physical quantities and their representation that motivates the creation of a negativity framework specific to physics.

\subsection{Initial Development}
\label{ssec:devNoNIP}

To create a physics-specific map of the natures of negativity, we began by generating a list of physics relationships involving an explicit negative sign (e.g., $\vec{F}_{\textrm{spring}} = -k\vec{r}$), and by considering base quantities (e.g., position, charge) and derived quantities (e.g., velocity, electric field) that can be negative or associated with a negative sign. Our intent was to develop a framework drawing on the practices and conventions of physics---which necessarily exist in a blended space of mathematics and physics---to make meaning of negativity.

Learning scientists have used card-sorting tasks to investigate mental organization of disciplinary knowledge \cite{chi1981,schoenfeld1982}. Experts are given cards showing various content with no pre-established groupings. They are then asked to sort the cards into groups that they feel make the most sense and describe each group. Two of the authors employed a modified card-sorting task of the quantities and relationships that make up the introductory physics course. The sorting process was modified to focus specifically on the role of the negative sign for each quantity or process. The task resulted in broad categorization based on physical similarities, gradually refined through further discussion and comparison with other quantities. Creation of a ``Change'' category exemplifies this process. The importance in physics of change and of conserved quantities led us to categorize many different and seemingly disparate uses of the negative sign in a category related to change, because of an underlying connection to the calculation of and reasoning about change. This includes the negative sign as an operator (to calculate change or to signify the physical removal of a quantity from a system); the negative sign as an indication of a decrease in a quantity; and the compound use of the negative for signifying a change in a quantity as well as calculating it. 

Along with \textit{Change}, the emergent categories were \textit{Direction} and \textit{Opposition}. A fourth category, \textit{Compound} was added to account for cases that require interpretation of multiple negative signs in a single context. We note that the \textit{Direction} and \textit{Opposition} categories are supported by the categories isolated Chiu's study  \cite{chiu2001}. Phenomena that arise due to the parallel or antiparallel orientations of two quantities are ubiquitous throughout physics (e.g., speeding up/slowing down, friction and air resistance, electromagnetic induction). Direction and Opposition are central natures of signed quantities in physics, and hallmarks of physics reasoning.

To allow for further refinement, subcategories emerged within each of the main categories.
Some subcategories also specify the mathematical function of the negative sign (as in the \textit{Change} category's subcategory \textbf{Difference (operator)}). Table \ref{tab:NoNIP} shows the resulting map of the natures of negativity in introductory physics. We do not attempt to further explain the meanings of the categories and subcategories shown in the table, as further research led us to reorganize this early version of the NoNIP.
\begin{table*}
\footnotesize
\caption{Initial version of the natures of negativity in introductory physics, a framework for the different uses of the negative sign in introductory physics}
\label{tab:NoNIP}

\begin{ruledtabular}
\begin{tabular}{ c  c  c  c }
(D) Direction & (O) Opposition &(Ch) Change & (Co) Compound \\ \hline
\multicolumn{1}{l}{\textbf{1. Location}}  	 	& \multicolumn{1}{l}{\textbf{1. Opposite type}}	 	&  \multicolumn{1}{l}{\textbf{1. Removal (operator)}}  &\multicolumn{1}{l}{\textbf{1. Scalar rates of change}} \\
$x$	& \textit{Q (charge)}&  $0 - (-5 \mu C)$& $\frac{\textrm{d}\phi}{\textrm{d}t}$ \\
\multicolumn{1}{l}{\textbf{2. Direction of motion}}	& \multicolumn{1}{l}{\textbf{2. Opposes}}				&  \multicolumn{1}{l }{\textbf{2. Difference (operator)}}& \multicolumn{1}{l}{\textbf{2. Base + change}} \\
$v_x, \Delta x$	& $\vec{F}_{12} = - \vec{F}_{21}$& $E_f -E_i$ & $\phi + \frac{\textrm{d}\phi}{\textrm{d}t}t$ \\
	$p_x$& $\vec{F} = -\vec{\nabla}U$	& $\vec{p}_f -\vec{p}_i $  		& $\vec{v}+\vec{a}t$ \\
\multicolumn{1}{l }{\textbf{3. Other vec. quant. comp.}} & $\mathscr{E} = -\frac{\textrm{d}\Phi_B}{\textrm{d}t}$		&
\multicolumn{1}{l }{\textbf{3. System scalar quantities}}    & \multicolumn{1}{l}{\textbf{3. Products $f(x)dx$}} \\
$E_x, B_x$ & $\vec{F}=-k\vec{r}$ & $\Delta K, \Delta E$ & $E(r)dr$ \\
$F_x, L_z$	& \multicolumn{1}{l}{\textbf{3. Scalar products}}			& $\Delta S$ 	&  $P(V)dV$	\\
	$a_x$ & $W = \vec{F} \cdot \Delta\vec{x}$ & \multicolumn{1}{l}{\textbf{4. Scalar, vector change}}& \multicolumn{1}{l}{\textbf{4. Models}} \\ 
$\Delta p_x, \Delta v_x$ & $\Phi =\vec{B}\cdot\vec{A}$ &	
$\Delta E=E_f-E_i, \Delta V=V_f-V_i$ & $W_{net,ext}=\Delta E$  \\
\multicolumn{1}{l}{\textbf{4. Above/below reference}}	&	&	$\Delta\vec{p} = \vec{p}_f -\vec{p}_i $  
& $\vec{F}_{net} = m\vec{a}$ \\
\textit{T (temperature)} &  &  & $\Delta U = Q - W$ \\
\textit{V (electric potential)}	& 	&	&  
\end{tabular}
\end{ruledtabular}
\end{table*}

The initial version of the NoNIP, Table \ref{tab:NoNIP}, was extensively validated. We assessed face validity by surveying introductory physics textbooks, using NoNIP to categorize all instances of the use of negative signs. With one notable exception (negative exponents), we found that all uses of the negative sign could be categorized satisfactorily using the NoNIP. 

Expert validation of the NoNIP was conducted by performing formal, semi-structured interviews with two experts each in mathematics and physics. Our physics experts are experienced introductory-level instructors; our mathematics experts taught introductory- and intermediate-level undergraduate mathematics (for example, single- and multivariable calculus and differential equations), with sufficient background in physics to understand physics-specific meanings (e.g., one of our mathematics experts had an undergraduate degree in physics). During interviews lasting 30--60 minutes, experts were asked to comment on the appropriateness of the map in the context of introductory physics, as well as any uses of the negative sign that were not compatible with the NoNIP framework. Mathematics experts were also asked to comment on our interpretations of the negative sign from a more algebraic perspective. Expert comments were very supportive. 

\subsection{Steady state version of the NoNIP}
\label{ssec:currentVersionNoNIP}

While feedback from experts in interviews---and during less formal interactions during conference presentations---was positive, comments led to a number of changes to the NoNIP. These changes were initially small, but ultimately led to a re-organization. Each of the mathematics experts interviewed stressed the importance of recognizing the meaning of ``zero'' in any given context. This led us to consider the negative sign in the ``unary'' sense (that is, attached to a single quantity) separately from other uses, and allowed us to distinguish more clearly between scalar and vector quantities. Comments by physics experts led us to consider when quantities are \emph{opposite} one another (as $-5$ is opposite $+5$, such that $+5 + (-5) = 0$) and when quantities \emph{oppose} one another (as in Faraday's law, $\mathscr{E}=-\frac{\textrm{d}\Phi_B}{\textrm{d}t}$). Finally, we recognized the importance of differentiating between the negative sign as an operator and its other uses. 

Although we reorganized the NoNIP to have categories that are more algebraically based, we believe that the major strengths of both the initial and revised versions is the physical interpretation present in the main categories as well as the subcategories.

The revised NoNIP is shown below in Table \ref{tab:newNoNIP}. Here, we focus on the functions of the negative sign: specifying its use with a \emph{quantity}, defining the \emph{relationship} between two quantities, or as an \emph{operation}. The subcategories give meaning specific to physics. We have removed the \emph{compound} category because it is not parallel to the others, which will be discussed more fully below. It is worth noting that some quantities (such as mechanical work and electric charge) appear in multiple categories. This speaks to the challenges that students face when trying to decode and make sense of negative signs in introductory physics.

\begin{table*}
\caption{Current version of the natures of negativity in introductory physics}
\label{tab:newNoNIP}
\begin{tabular*}{\textwidth}{ c@{\extracolsep{\fill}}c  c }
\hline \hline
(Q) Quantity	& (R) Relationship  & (O) Operation \\ \hline
\textbf{1. Scalar}  & \textbf{1. Opposes}	&  \textbf{1. Removal (physical)} \\
\multicolumn{1}{l}{a. Type (charge only)}	& 	\multicolumn{1}{l}{a. Scalar} 	&  $0 - (-5 \mu \textrm{C})$, $m_{\textrm{total}}-m_\textrm{a}$	 		\\
\multicolumn{1}{l}{b. Change/rate of change}	& 	$\mathscr{E} = -\frac{\textrm{d}\Phi_B}{\textrm{d}t}$	&  \textbf{2. Difference (temporal)} \\
 $\Delta E, dV, \frac{\textrm{d}\phi}{\textrm{d}t}$ & \multicolumn{1}{l}{b. Vector}	& $E_{\textrm{f}} -E_{\textrm{i}}$	 \\
 \multicolumn{1}{l}{c. Comparison to reference}  	& 
$\vec{F} = -\vec{\nabla}U$	& $\vec{p}_{\textrm{f}} -\vec{p}_{\textrm{i}} $ \\
$T,V,E,t$   & $\vec{F}_{\textrm{spring}}=-k\vec{r}$ & \textbf{3. Difference (other)}\\
\multicolumn{1}{l}{d. Models/Convention} & 	\textbf{2. Opposite}& 1st Law of Thermodynamics ($Q-W$)\\
$W_{\textrm{net,ext}}$, Heat $(Q)$, Current $(i)$ & $+5~\mu\textrm{C} + (-5\mu\textrm{C})$ $= 0$	 & Pathlength difference ($\Delta D$)\\
\textbf{2. Vector} 	& $\vec{F}_{12} = - \vec{F}_{21}$ & 
Distance from equilibrium ($\Delta x$) \\
\multicolumn{1}{l}{a. Direction from origin} & 	
\textbf{3. Relative Orientation}	&  	Electric potential difference ($\Delta V$)	\\
$x, E_{x}, v_{x}$ &  $\vec{F} \cdot \Delta\vec{x}$, $\vec{E} \cdot \vec{A}$	&
\textbf{4. Removal (modeling)}	\\
\multicolumn{1}{l}{b. Direction of change} & \textbf{4. Negative exponents} & $I_{\textrm{net}}=I_{\textrm{disk}}-I_{\textrm{hole}}$	\\
$\Delta p_{x}, \Delta v_{x}$ & $e^{-\frac{t}{\tau}}$, $r^{-2}$ & \\
		 \hline \hline
\end{tabular*}
\end{table*}

\subsubsection{Description of the natures of negativity in introductory physics}
The \emph{Quantity (Q)} category is most similar to the mathematical \emph{Unary} category described by Vlassis (i.e., identifies a number as negative), with the negative sign attached to a single quantity. The Q category is subdivided for scalar (1) and vector (2) quantities; because the negative sign associated with a quantity has different meanings for scalars and vectors, this subdivision is appropriate here.

Four subcategories exist for scalar quantities. The first subcategory, \textbf{Type} (1a) is reserved for electric charge. In the case of electric charge, sign specifies the type of charge. We can identify no other contexts that use the negative sign this way. The subcategory \textbf{Change} (1b) is for scalar quantities such as $\Delta E$, as well as scalar time rates of change, such as $\frac{\textrm{d}\phi}{\textrm{d}t}$, where a negative sign typically indicates that the quantity decreases with time. We also include differentials in this category, as a negative differential indicates an infinitesimal decrease in a quantity. Next, scalar quantities such as temperature, electric potential, and energy can be negative relative to an arbitrary reference point ``0''; these quantities are in the subcategory \textbf{Comparison to reference} (1c). Finally, we consider scalar quantities for which the sign carries important physical meaning that is an artifact (sometimes arbitrary) of given \textbf{Models/convention} (1d). We consider quantities such as heat $Q$ (which is negative for a system when heat is transferred out of that system), the net external work done on a system (which, when negative, signifies that the mechanical energy of the system decreases), and current $i$ (which is negative when opposite to the sense arbitrarily decided to be positive) to be described by this subcategory.

For vector quantities, we consider two subcategories for vector components: \textbf{Direction from origin} (2a), which describes the direction relative to a coordinate system or origin, and describes quantities such as components of position, electric field, and velocity; and \textbf{Direction of change} (2b) for components of differences of vector quantities, such as changes in momentum and in velocity. We consider these two vector subcategories to be distinct from each other. For isolated vector components such as $E_x$, we consider a single vector and how it compares to a coordinate system. For vector components such as $\Delta \vec{v}_x$, the direction tells us something about the difference between two vectors. 

Although the category \textit{Relationship (R)} is most similar to the algebraic category \textit{Symmetrical (operational signifier)} (i.e., when a negative sign is used to take the opposite of or invert a number or operation), we do not consider the negative sign to be indicative of an \emph{operation} for the quantities described in this category; rather, the negative signs in this category signify how quantities \emph{relate} to each other. While in mathematics a negative sign may be used as an operator that defines one quantity as another's opposite, in physics the negative sign may describe that two quantities are inversely related. In this category (unlike in the Quantity category) we did not immediately sub-divide relationships by their scalar or vector nature. As stated above, the meaning of the negative sign is distinctly different for scalar and vector quantities. For use in relationships with explicit negative signs, however, the meaning of the negative sign is very similar for many scalar and vector relationships.

Our first subdivision is for relationships in which quantities \textbf{Oppose} each other. The relationships in this subdivision have an explicit negative sign, as the negative sign signifies the opposing nature of the relationship between quantities. There are two subcategories in this division. In the \textbf{Scalar} (1a) category, we have relationships such as that described by Faraday's Law, where the EMF opposes the time rate of change of the magnetic flux. Similarly, in the \textbf{Vector} (1b) category, we have vector relationships such as Hooke's Law, where the force exerted by a spring opposes the spring's displacement. 

In the next subdivision, we consider quantities that are \textbf{Opposite} to each other. These are not quantities that are inversely related by a physical relationship such as Faraday's or Hooke's Law. Despite this, we put such quantities in the Relationship nature, as the negative sign still indicates that these quantities are opposite to \emph{another quantity}. Examples include positive and negative charge, and the members of a Newton's Third-Law Force Pair \footnote{It was only after much deliberation that Newton's Third-Law Force Pairs were categorized as ``opposite'' and \emph{not} as ``opposes.'' Although it would be reductive to characterize all of the relationships in the ``opposes'' category as \emph{causal}, many of those relationships are at least \emph{directional}. Members of a Third-Law Force Pair are not causal or directional in that fashion, despite the convention of calling them ``action-reaction'' forces. Further, the forces in a Third-Law Force Pair act, by definition, on separate objects.}. 

We include \textbf{Relative Orientation} for scalar products such as those used to calculate mechanical work ($\vec{F}\cdot \Delta \vec{x}$) and electric flux ($\vec{B}\cdot\vec{A}$). A negative scalar product indicates that the factor vectors have components that are oppositely-oriented. We include scalar products in the relationship category because, as with other expressions and quantities in this category, a negative sign says something about how two quantities (in this case, the factor vectors) relate to each other. 

Finally, we added a subcategory for \textbf{Negative Exponents}. This is included in the Relationship category because negative exponents typically describe how one quantity relates to another: how a quantity decreases in time may be described by exponential decay, such as in circuits or damped harmonic oscillators; or how a quantity decreases in space is often described by $\frac{1}{r^2}$. Recognizing the equivalence between a quantity with a negative exponent and the inverse of that quantity with a positive exponent is vital and (in our experience) nontrivial. This addresses a major shortcoming that existed in the original version of the NoNIP, which did not allow for straightfoward categorization of negative exponents.

Our third base category is \textit{Operation (O)}, for instances when the negative sign is used to perform the mathematical function of subtraction. This category is similar to the algebraically-based category ``Binary'' described by Vlassis, which also describes uses of the negative sign to indicate subtraction. There are four subcategories: 1) \textbf{Removal}; 2) \textbf{Difference (change)}; 3) \textbf{Difference (other)}; and 4) \textbf{Removal (modeling)}. The \textbf{Removal} category is for uses of the negative that signify a physical removal of some quantity (such as electric charge or mass) from a system, whereas the \textbf{Difference (change)} subcategory is used for calculating the temporal change in a quantity, such as $E_{\textrm{f}} -E_{\textrm{i}}$ or $\vec{p}_{\textrm{f}} -\vec{p}_{\textrm{i}}$. The third subcategory is for the operation used to calculate differences that are not necessarily temporal in nature. An example is for the calculation of displacement for Hooke's law, with $\Delta \vec{x} = \vec{x}_{\textrm{displacement}}- \vec{x}_{\textrm{equilibrium}}$. Here, we aren't considering $\Delta \vec{x} = \vec{x}_{\textrm{f}}- \vec{x}_{\textrm{i}}$, only the displacement from equilibrium. Similarly, pathlength difference $\Delta D$ can be determined by subtracting one pathlength from another. Non-$\Delta$ differences, such as the subtraction of work done by a system from the heat supplied to the system, as in the First Law of Thermodynamics, are also included in this subcategory. The fourth subcategory (\textbf{Modeling (removal)}) is reserved for the non-physical removal of one quantity from another for the purposes of modeling a more complex situation. This type of subtraction is typified by the calculation of the moment of inertia of a solid such as a disk with a hole.




\subsubsection{Compounding multiple natures of the negative sign}
Expert feedback led us to recognize that creating a \emph{Compound} category in the original NoNIP somewhat hid the cognitive difficulty students encounter unpacking the multiple natures of negativity from a single expression or equation. Because of the many possible combinations of multiple signed quantities appearing in an expression, each compound context poses a unique challenge. It is in these compound contexts that even very strong students struggle most. A significant difference between the revised version of the NoNIP and the original version shown in Table \ref{tab:NoNIP} is the new version's lack of a Compound category. We believe that Quantity, Relationship, and Operation represent the individual natures of negativity in introductory physics thoroughly. The task of combining them in compound cases is an example of the sophisticated reasoning with familiar mathematics that is characteristic of well-developed quantitative literacy.

\begin{figure}
\begin{center}
\includegraphics[width=.45\textwidth]{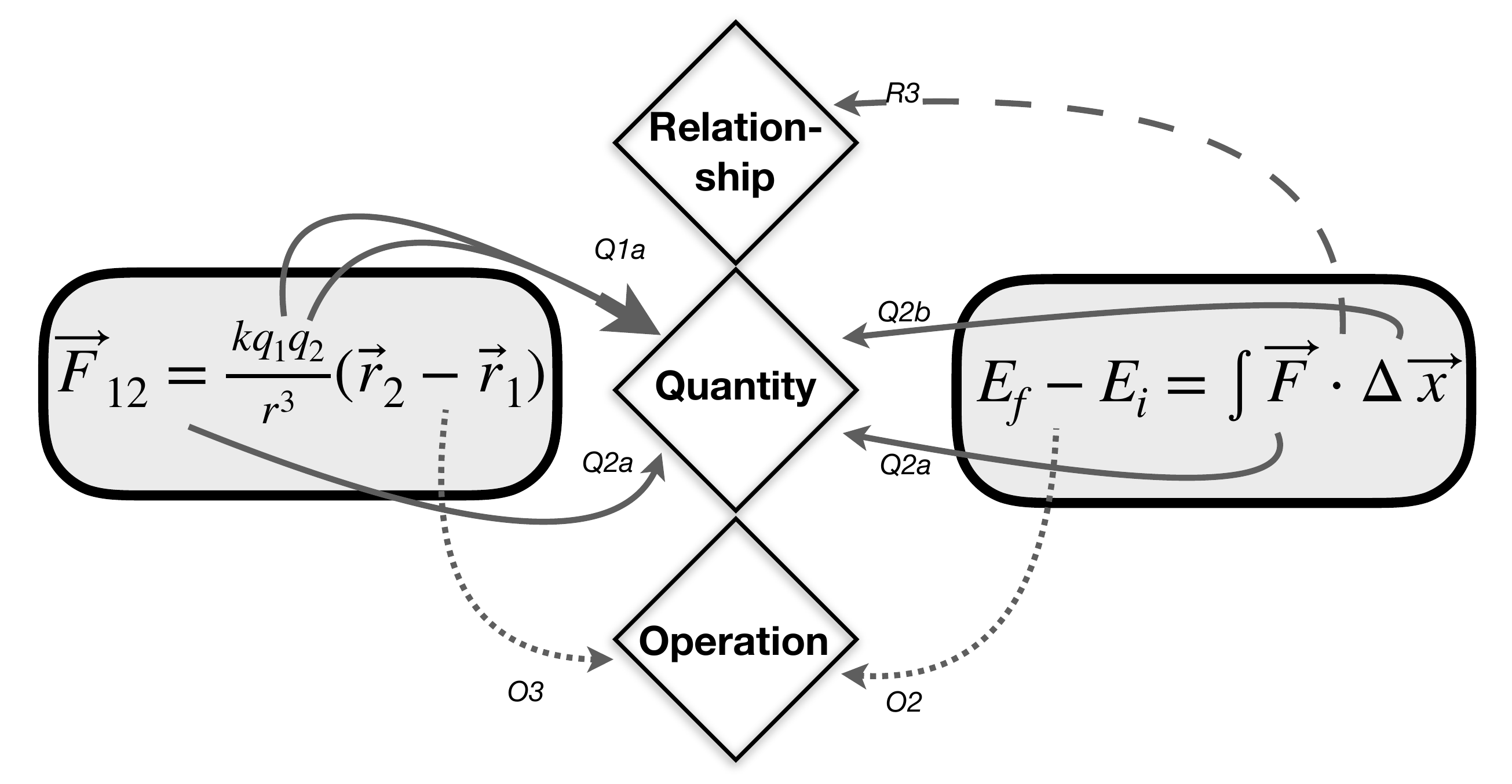}
\caption{Compound use of negative signs typical of introductory physics: Coulomb's law, the force of object 1 on object 2 (left), and the work-energy theorem (right).}
\label{fig:CompoundDiagram}
\end{center}
\end{figure}

Many models in physics involve more than one unique nature of negative quantities; we consider the compound expressions ubiquitous in physics to be incomprehensible without flexibility between the three main natures of negativity. Figure \ref{fig:CompoundDiagram} maps two examples to the current NoNIP, which reveal various natures of negativity from all three categories that must be understood in order to make physical sense of central ideas in introductory physics. One must note that some quantities (such as electric charge) appear under multiple natures, as the sign carries multiple meanings. Moreover, in some cases, such as Coulomb's Law, multiple negative signs may ``cancel out.'' Such ``hidden'' negative signs make keeping track of the signs of individual components more challenging. When Coulomb's Law is used with two charges with \emph{opposite} signs, students must also make sense of what the resulting negative sign implies. For physics experts, a negative sign implies that the force between the two charges is attractive, but this meaning is also described by the mathematics: that $\vec{F}_{\textrm{on 1 by 2}}$ is oppositely directed to $\vec{r}=\vec{r}_{\textrm{1}}-\vec{r}_{\textrm{2}}$. We contend that this physical interpretation of a quantified relationship is a hallmark of expert-like reasoning.

The identification of multiple meanings of negative signs in a single context is challenging and the ability to do so is associated with expert-like thinking. It should not be considered to be a skill at the same level as identifying the meaning in less complex cases; rather, it should be viewed as a culmination of reasoning with and about sign. We argue that flexibility between the various natures that appear in NoNIP may better prepare students for the challenge of combining them into a single equation. Indeed, much of the research of more advanced student reasoning about sign has been in the context of compounded use of the negative sign. In the following section, we describe such research, using the NoNIP as a framework by which student difficulties can be understood. In doing so, we provide examples of how to use the 3-nature NoNIP to categorize compound quantities and relationships.

Like the original version of the NoNIP, this version has been validated using reviews of introductory-level physics textbooks as well as formal interviews with mathematics and physics experts. The validation interviews were especially useful in refining the characterization of the negative sign in introductory physics, leading to the creation of new subcategories as well as reclassification of some quantities and relationships. We consider the current version to be in a steady state.

\section{Applying the framework}
\label{sec:Applications}
In this section, we use the current, steady-state version of NoNIP as an analytical lens through which to view four recently published studies in physics and calculus that mostly involve advanced physics or math students. 

Bajracharya, Wemyss, and Thompson investigated upper-division student understanding of integration in the context of definite integrals commonly found in introductory physics, but with all physics context stripped from the representation. Specifically, the variables typically used in physics contexts were replaced with $x$ and $f(x)$ \cite{bajracharya2012}. Their results suggest difficulties with the criteria that determine the sign of a definite integral.  Students struggle with the concept of a negative area-under-the-curve, and in particular negative directions of single-variable integration. In research related to student understanding of integration and negativity, Sealey and Thompson interviewed undergraduate and graduate students to uncover how they made sense of a negative definite integral. Undergraduate (beyond introductory) and graduate mathematics students had difficulty making meaning of a negative differential in the context of integration \cite{Sealey2016}. The struggles these researchers described can be seen through the lens of NoNIP as struggle with the product of the integrand, $f(x)$ (generic Q in NoNIP), and the differential, $\textrm{d}x$ (Q1.b in NoNIP), each of which can independently be negative. Making meaning of the negativity of the integrand (generic Q in NoNIP) was less of a struggle for the students in these studies than was the notion of a negative differential (Q.1b in NoNIP), which has application throughout physics. The researchers report that ``none of the students thought about $\textrm{d}x$ as a signed quantity on their own accord, but with prompting from the interviewers, some were able to do so.'' Encountering the differential as a small change in quantity, and being provided opportunity to think about it in this way, provides a context that has been shown to help \cite{roundy2014}. NoNIP can support this kind of explicit reasoning about sign in this important context. 

A study conducted by Hayes and Wittmann situated in the context of sophomore-level mechanics investigates the negative signs and quantities associated with the equation of motion of an object thrown downward, with non-negligible air resistance \cite{Hayes2010}. The interviewed student struggles with treating one-dimensional acceleration as a signed quantity, and feels there should be an additional negative sign included to indicate that the acceleration is ``negative,'' or opposing the motion. The authors explain student difficulties with negativity creating a notion of implicit and explicit ``minus'' signs. They conclude that the multiple natures of the negative sign are a source of cognitive conflict that manifests as sensemaking about ``outer and inner minus.'' An ``outer minus'' is a negative sign that is assigned, for example, by choice of coordinate system; an ``inner minus'' is one that is associated with variables that may be negative---buried in the physics meaning. Through the lens of NoNIP, ``minus'' is an operator, and negative signs are used to represent many mathematical objects and relationships in physics. In this particular context of one-dimensional motion, both the (one-dimensional) acceleration and initial velocity have ``explcit'' signs that carry physical meaning. In the framework of NoNIP, the student struggles with Q.2a in the contexts of one-dimensional acceleration and velocity. The negative sign that modifies the $cv$ term is used as R.1b, 
to indicate that the force is in the opposite direction to the velocity. In the process of combining these terms, the student struggles 
to make sense of the equation of motion. The cognitive load of negativity associated with the individual terms contribute to a higher-level struggle of making physical sense. The burden of these compound natures of the negative sign resulted in roughly half of this group of physics majors feeling it was necessary to include an extra negative sign in the equation of motion so that the mathematics would match the physics.

In their study of negativity in junior-level Electricity and Magnetism, Huynh and Sayre describe the in-the-moment thinking of a student solving for the electric field due to two equal and opposite charges along the axis that passes through them \cite{huynh2018}. The authors focus on student reasoning about the sign of the electric field vector component along the axis of symmetry in three regions of space---to the left of one charge, between the two charges and to the right of the other charge. They report that four students solve this problem in an oral exam, and none get the directions correct on their first attempt. 

    The solution involves an algebraic superposition of the field due to each charge individually. The authors detail one representative student's development of an increasingly blended approach that is situated in a mental space informed by both mathematical and physical concepts. The student starts reasoning about the direction of the field using Coulomb's law by (unintentionally) combining multiple natures of negativity into one. In Coulomb's law, the signs combine multiplicatively from two sources:  the direction of the displacement vector, $\vec{r}_2-\vec{r}_1$, and the sign of the source charge (see Fig.\ \ref{fig:CompoundDiagram}). 
The student first uses the canceling feature that multiplying two negative numbers always results in a positive number, without explicitly considering the source of each negative sign, and then reflects based on physics considerations why that approach doesn't make sense.

The student considers $\hat{x}$ to be a proxy for $\hat{r}$, without considering that $\hat{r}$ is connected to the physics (the difference of the two position vectors with respect to an origin) while $\hat{x}$ is a feature of the coordinate system. By making this substitution, the student glosses over an important source of negativity in the final solution and it keeps him from being able to calculate the direction of the electric field that he predicts using physics principles. His conceptual understanding of the physics is strong, but he can't make his calculation match because he isn't considering that the source of the negative signs have deep physical meaning beyond the charges involved. Seen through the lens of NoNIP we can see evidence of the student first conflating the natures Q.2b with Q.2a, not recognizing that they are the same thing. The authors summarize the student's confusion regarding the relative signs of the contributions to the electric field due to each charge individually:

\begin{quote}
    This result clearly conflicts with their relative direction because he has double associated their opposite direction with inappropriate application of destructiveness. [The student] tries hard to determine where another negative sign could come from, such as the denominator, to cancel one negative sign for the whole term. Finally, he decides to absorb the destructive meaning of the sign into the opposite-direction meaning of the electric field vector and changes the second negative sign of the whole term back into a plus sign, which supports the fact that they are in opposite directions. However, [the student] has not considered the sign commensurate with the relative direction of $\hat{E}$ and $\hat{x}$, leading to his solution having the opposite sign to the correct answer \cite{huynh2018}.
\end{quote}


Collapsing the signs using arithmetic rules is a common approach first tried by the students in this study, which focuses on the multiplicative rules of signed numbers rather than the physics of the meaning of the signs. Next, the student rarefies his approach as he considers more carefully the natures of negativity in the context of the problem.  The student reflects ``\dots I should have figured it out \dots which direction it is. This is exactly what is changing signs, not necessarily the sign of the charge.'' After reconciling the basic level, then he struggles with R.1b and its equivalence to Q.2b. The authors report that when the student moves on to the other two regions, which is essentially repeating the same reasoning sequence, the student encounters the same struggles but he is faster at obtaining a solution that matches his physical understanding of the system. The fact that he doesn't automatically and quickly solve the remaining two regions is evidence that this kind of reasoning is difficult. 

The authors conclude, and we agree, that the most sophisticated challenge occurs when the natures negativity are combined---the compound nature presents its own challenge in addition to the challenge associated with each nature individually. This case study reveals the cognitive difficulty when three natures of the negative sign must be made sense of in the context of a single equation, and illustrates the challenges associated with reasoning about the natures of negativity, even for strong majors. We believe it also reveals a hierarchy that lends plausibility to the NoNIP model being representative of emergent expert-like reasoning.

This section, in which previously published work is interpreted through the NoNIP framework, demonstrates the benefits of NoNIP: it allows comparisons across studies that involve different physics contexts and even different mathematical approaches. Both the Huynh \& Sayre and Hayes \& Wittmann studies involve upper-division topics and situations involving negative signs \cite{huynh2018,Hayes2010}; otherwise there are few similarities. The Hayes \& Wittmann paper is situated in classical mechanics and difficulties with ``inner and outer'' negative signs, while the Huynh \& Sayre paper investigates student ability setting up integrals in E{\&}M. By using the NoNIP framework, we can see that the two papers discuss similar aspects of the use of the negative sign. These similarities are much more specific than just using a negative sign in upper-division contexts. Further, the study by Bajracharya, Wemyss, Thompson \cite{bajracharya2012} is also about using negative sign in an upper-division context, but applying the NoNIP framework reveals a use of a different nature of negativity than the other two physics papers. We believe this illustrates the power of the NoNIP for researchers: identifying similarities across (and differences between) studies that are more than just superficial.

\section{Extension to signed quantities}
\label{sec:signedQuant}

Although we focus in this paper on our categorization of the negative sign in introductory physics, we recognize that students must make sense of the meaning of positive quantities and relationships as well. 
Our focus on the negative sign in this work is due to the assumption of ``positivity'' when a quantity has no explicit sign. While the practice of assuming an unsigned number has an implicit positive sign in front of it perhaps poses few problems understanding pure numbers, quantities in physics that aren't signed may be either unsigned scalars (e.g., speed, mass, time), or quantities in which the positive sign holds meaning (e.g., component of velocity, change in energy).

\begin{figure}
\framebox{\parbox{0.45\textwidth}{\raggedright

ME1: An object moves along the x-axis, and the acceleration is measured to be $a_x = +8 \textrm{ m/s}^2$. 

Consider the following statements about the ``+'' sign in ``$a_x = +8 \textrm{ m/s}^2$''. Pick the statement that best describes the information this positive sign conveys about the situation.
	\begin{enumerate}
	\setlength{\itemsep}{0.0ex}
	\renewcommand{\labelenumi}{\alph{enumi}.}
	\item The object moves in the positive direction
	\item The object is speeding up.
	\item \textbf{The object accelerates in the $+x$-direction}
	\item Both a and b
	\item Both b and c
	\end{enumerate}
    }}
\framebox{\parbox{0.45\textwidth}{\raggedright

ME2: A hand exerts a horizontal force on a block as the block moves on a frictionless, horizontal surface. For a particular interval of the motion, the work $W$ done by the hand is $W= -2.7\textrm{ J}$ . Consider the following statements about the ``$-$'' sign in the mathematical statement ``$W= -2.7\textrm{ J}$.'' The negative sign means:

	\begin{enumerate}
	\setlength{\itemsep}{0.0ex}
	\renewcommand{\labelenumi}{\Roman{enumi}.}
	\item the work done by the hand is in the negative direction
	\item the force exerted by the hand is in the negative direction
	\item the work done by the hand decreases the mechanical energy associated with the block
	\end{enumerate}
Which statements are true?
\begin{enumerate}
	\setlength{\itemsep}{0.0ex}
	\renewcommand{\labelenumi}{\alph{enumi}.}
	\item I only
	\item II only
	\item \textbf{III only}
	\item I and II only
	\item II and III only
	\end{enumerate}
    }}    
\framebox{\parbox{0.45\textwidth}{\raggedright

ME3: A cart is moving along the x-axis. At a specific instant, the cart is at position $x = -7$ m. Consider the following \\statements about the ``$-$'' sign in ``$x = -7$ m.'' Pick the statement that best describes the information this negative sign conveys about the situation.
	\begin{enumerate}
	\setlength{\itemsep}{0.0ex}
	\renewcommand{\labelenumi}{\alph{enumi}.}
	\item The cart moves in the negative direction
	\item \textbf{The cart is to the negative direction from the origin}
	\item The cart is slowing down
	\item Both a and b
	\item Both a and c
	\end{enumerate}
    }}    
\caption{Examples of a multiple-choice questions probing student understanding of signed quantities. ME1 is an example of a positive-quantity question, while ME2 and ME3 are negative-quantity questions.}
\label{fig:SignedQuant}
\end{figure}

\begin{figure}
\begin{center}
\includegraphics[width=.40\textwidth]{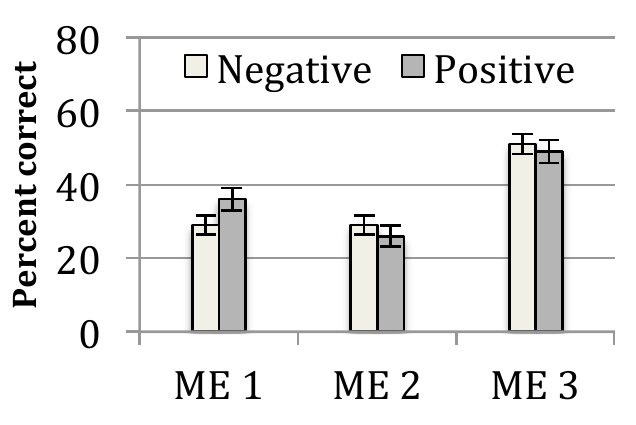}
\caption{Percentage of students who answered correctly for positive and negative versions of mechanics items, $N_{\textrm{pos}}=242$, $N_{\textrm{neg}}=309$; the error bars represent the binary standard error.}
\label{fig:negPosResults}
\end{center}
\end{figure}

To investigate student understanding of signed quantities more generally, we administered three questions about positive or negative quantities in a multiple-choice format, shown in Table \ref{fig:SignedQuant}, to students enrolled in a calculus-based introductory physics course at a large, diverse, public R1 university. 
Each student received either all three negative or all three positive versions. Figure \ref{fig:negPosResults} shows the results from the negative and positive versions of the mechanics items ($N_{\textrm{pos}}=242$, $N_{\textrm{neg}}=309$). 

To determine the effect size, we use the odds ratio. It is a useful effect size measure that describes the likelihood of an outcome occurring in the treatment group compared with the likelihood of the outcome occurring in the control group by forming a ratio of the two. An odds ratio of $1$ would indicate that the odds are exactly the same. If we consider the hypothesis that positive quantities pose fewer challenges for students than negative ones, we can consider the positive questions as the control and the negative questions as the treatment. 

For items ME2 and ME3, effect sizes determined from odds ratios are $1.2$ and $1.1$ respectively, which imply that students struggle with both positive and negative versions with roughly equal likelihood on these questions. The effect size for item ME1 is $0.74$, which is a statistically small effect size indicating that students find the negative version of this question slightly more difficult than the positive version \cite{maher2013}. Our experience as instructors has led us to recognize that students tend to inappropriately associate negative (positive) acceleration with decreasing (increasing) speed regardless of the coordinate system. We suspect that the small difference in student performance on the two versions of item ME1 are related to this difficulty, and thus we do not interpret the finding as evidence that students experience inherent difficulty with negatively-signed quantities (i.e., relative to positive quantities).

These results indicate that students have difficulty interpreting the meaning of the sign of a quantity, \textit{regardless of the sign}; students may not recognize that the sign specifies the direction of a vector component relative to a coordinate system, or that the sign of a scalar quantity such as work indicates how the energy of a system changes. 

Informal conversations with students also indicate that students sometimes fail to see the significance of the \emph{relationship} between two quantities that are positively correlated (e.g., that Newton's Second Law tells us that the acceleration of a system is always in the same direction as the net force exerted on that system). Such an understanding is crucial not only for nominally causal relationships, but also for understanding feedback loops and accumulated change. While interpreting negative signs is a pressing issue in physics learning because negative signs are explicitly used, these difficulties fall under the umbrella of student difficulties with the interpretation of signed quantities in general---which has not been studied in depth.

\section{Conclusion}
\label{sec:Conclusion}
Negative signs in physics have nuanced and varied interpretations that can pose a challenge, even to majors. In this paper, we present the NoNIP framework for categorizing expert uses of the negative sign which has undergone expert validation and revision; it is presented in its steady state. We anticipate that the NoNIP table can be useful across the physics education research community. For researchers, the NoNIP table can serve as a map of the natures of negativity, and a starting point for thinking about sign, as part of the broader context of mathematical reasoning development in physics context. We have demonstrated how viewing published research through the NoNIP framework can help bring out patterns between the findings that weren't clear before.  

We present evidence that students struggle to make sense of positive signs as much as they do negative signs. Student difficulties with positivity aren't as noticeable because in practice experts assume the absence of sign means the quantity is positive, so, unlike a negative quantity, there is no symbol there to decode. We intend for the research presented in this paper to extend to signed quantities more generally, and increase awareness that students will benefit from making sense of the meaning of positive quantities as well as negative. The sign of a quantity, along with the magnitude of the quantity and its units, are part of what defines a quantity and how we understand it in most physics contexts \cite{Brahmia2019}. 

As a tool for instructional development, NoNIP can benefit both curriculum developers---who can use the NoNIP table to help guide their efforts to situate signed quantity reasoning in the broader context of the materials that they develop---and instructors.  We close by presenting some recommendations for instruction that can inform instructors about the organization of their expertise, which can thus influence how they talk about and present material to the novices in their classroom. Developing this awareness can help students become more cognizant of the natures of both positivity and negativity in physics. Acknowledging the nuances, rather than assuming the mathematics to be trivial, can create access for students that otherwise might not exist. We offer three suggestions as a start, fully anticipating that expert instructors will devise their own ways also:
\begin{enumerate}
\item Quantities that are inherently signed quantities should be prefaced with a negative sign when the quantity is negative, and a positive sign when the quantity is positive, e.g., $x_{\textrm{o}}=+ 40$ m. Priming students to expect that real-world quantities often have signs associated with them that carry meaning, and that ``no sign'' is a different kind of quantity than a positively signed quantity, can help establish a physics habit of mind that the sign carries scientific meaning, and eventually that vector quantities have different mathematical properties than scalar ones.

\item Orientation (along a particular axis) and sense (positive or negative) are not always explicit in coordinate systems. In problems associated with motion, aligning the positive coordinate axis with the direction of motion eliminates the need for signed quantities when discussing velocity. This choice, however, could be a missed opportunity to distinguish between orientation and sense. The opposite coordinate choice can prime students to consider the signed nature of position, velocity, and subsequent vector quantities they encounter.

\item Sign and operation are often conflated using an equals sign (e.g., $5 + (-3) = 5 - 3$), and unsigned numbers are assumed positive. Adding a negative quantity and subtracting a positive one often have different meanings in physics contexts (e.g., adding electrons). Although these operations yield the same arithmetic results, conflating them may lead students to struggle with the distinctions between sign and operation. We suggest using the term ``minus'' for the operation of subtraction, and the term ``negative sign'' to describe the symbol.
\end{enumerate}

In addition to enriching subsequent physics learning, a focus on natures of sign in physics contexts can also enrich the corequisite mathematics learning. Sealey and Thompson report on a context in which physics helps math students make sense of negativity in calculus. The researchers observed that invoking a physics example of a stretched spring helped catalyze sensemaking---the physics helped them to make sense of an abstract binary nature of the negative sign \cite{Sealey2016}. We suggest that there is symbiotic cognition possible in which both mathematics and physics learning can be enriched by conceptualization of the other, and that reasoning about sign provides a rich context. We present NoNIP as a representation of the natures of negativity providing a step in that direction.

\begin{acknowledgments}
We thank Peter Shaffer for his support with data collection that informed this work,  and the entire Physics Education Group at the University of Washington for vibrant discussion and feedback. Brian Stephanik's feedback and physics content expertise was also instrumental in the construction of the NoNIP. We thank Roy Montalvo at Rutgers University for sharing his creative software innovation that helped to make the data collection run smoothly. We also thank Laurie Smith for fruitful conversations regarding negative exponents.
\end{acknowledgments}

\bibliography{NoNIP}

\end{document}